\newcommand{\subI}{_{\rm _I}}
\newcommand{\subR}{_{\rm _R}}
\newcommand{\subs}[1]{{\rm _{s#1}}}
\newcommand{\sube}[1]{{\rm _{e#1}}}
\newcommand{\subi}[1]{{\rm _{i#1}}}
\newcommand{\dR}{\delta\subR}
\newcommand{\dI}{\delta\subI}

\newcommand{\vff}{v_{\rm ff}}

\newcommand{\rhoa}{\rho_{\rm a}}

\newcommand{\vb}{{\bf v}}

\newcommand{\tL}{{\tilde \Lambda}}
\newcommand{\psic}{\psi_{\rm c}}
\newcommand{\psiei}{\psi_{\rm ei}}

\newcommand{\kb}{k_{_{\rm B}}}

\newcommand{\GS}{\lower.5ex\hbox{$\buildrel>\over\sim$}}
\newcommand{\LS}{\lower.5ex\hbox{$\buildrel<\over\sim$}}

\newcommand{\etal}{et al.\ }
\newcommand{\eg}{e.g.\ }
\newcommand{\ie}{i.e.\ }

\documentstyle[12pt,epsfig]{article}
\def\reference{\parskip 0pt\par\noindent\hangindent 0.5 truecm}

\begin{document}
\title{Effects of lower boundary conditions on the stability
of radiative shocks}

\author{Curtis J. Saxton}

\date{}
\maketitle

{\center
Department of Physics \& Theoretical Physics, Faculty of Science and\\
Research School of Astronomy \& Astrophysics,\\
Australian National University, Canberra ACT 0200
\\saxton@mso.anu.edu.au\\[3mm]
}

\begin{abstract}
Thermal instabilities can cause a radiative shock to oscillate,
thereby modulating the emission from the post-shock region.
The mode frequencies are approximately quantised in analogy to
those of a vibrating pipe.
The stability properties depend on
the cooling processes, the electron-ion energy exchange
and the boundary conditions.
This paper considers the effects of the lower boundary condition
on the post-shock flow,
both ideally and for some specific physical models.
Specific cases include
constant perturbed velocity, pressure, density, flow rate, or temperature
at the lower boundary,
and the situation with nonzero stationary flow velocity at the lower boundary.
It is found that
for cases with zero terminal stationary velocity,
the stability properties are insensitive to
the perturbed hydrodynamic variables at the lower boundary.
The luminosity responses are generally dependent on
the lower boundary condition.
\end{abstract}


\bigskip

\section{Introduction}

A system that depends on the length- or time-scales of radiative cooling
or other energy exchange processes
may be thermally unstable (Field 1965)
if perturbations from the initial or equilibrium condition
alter the cooling scale in a manner that enhances the perturbation.
The effects of radiative cooling are expressed in terms of 
a function of local hydrodynamic variables such as density and temperature.
Qualitatively, a thermally unstable system has a form of cooling function
which increases the cooling length when the shock temperature increases
and conversely the cooling length decreases
when the shock temperature lowers.
Langer, Chanmugam \& Shaviv (1981)
discovered such a thermal instability in
the radiative accretion shocks of white dwarfs
(see \eg reviews by Cropper 1990, Wu 2000)
This instability was verified by subsequent studies
using different numerical techniques
(\eg Imamura, Wolff \& Durisen 1984),
and linear stability analyses
(\eg Chevalier \& Imamura 1982).

Linear analyses
(Chevalier \& Imamura 1982; Imamura \etal 1996;
Saxton \etal 1998; and Saxton \& Wu 1999)
indicate that the shock oscillations have a sequence of eigenmodes.
If bremsstrahlung dominates the cooling
then there tends to be a fundamental mode stable against oscillations,
and unstable overtones.
Depending on its relative efficiency,
cyclotron cooling tends to stabilise modes,
except when the electron-ion energy exchange is comparatively inefficient,
in which case modes can be destabilised with increasing cyclotron
efficiency.

T\'{o}th \& Draine (1993) found that
when bremsstrahlung cooling is the dominant cooling process
then the frequencies resemble those of a pipe open at one end,
(oscillation frequency $f\propto n-\frac12$ for integer harmonic number $n$).
This pattern persists for two-temperature shocks
when the electron-ion exchange is efficient
(Imamura \etal 1996, Saxton \& Wu 1999, Saxton 1999)
When cyclotron cooling is very efficient,
so that the electron-ion exchange process is unable to maintain
strong thermal coupling between electrons and ions,
the modes are more like those of a doubly-closed or doubly-open pipe
$f\propto n$,
(Saxton \& Wu 1999).

The standard analytic treatment defines the lower boundary
as the place where the flow velocity becomes zero in the
time-independent solution
and does not oscillate in the time-dependent response to perturbation
(\eg accreted material diffuses across a stellar surface very slowly
compared to the oscillatory timescale). 
The pressure, density and other hydrodynamic variables 
are nonzero and are allowed to oscillate. 
Different numerical investigations 
treated the lower boundary differently:
\eg dense matter near the lower boundary was deleted in Imamura (1985);
was merged with neighbouring cells in Imamura, Wolff \& Durisen (1984);
and may have had its gradients softened by the spatial filter on the 
Eulerian grid of Langer, Chanmugam \& Shaviv (1981).
These diverse treatments produced some different results.
Langar \etal (1981, 1982) find a fundamental mode that is unstable to
oscillations;
whereas other studies 
(\eg Imamura 1985; 
Imamura \etal 1984;
Wolff \etal 1991; 
Wood \etal 1992)
found that the first unstable mode is the first overtone,
and the fundamental mode is damped,
consistently with stability analyses.
The collaborations also produce different results
regarding the phasing of oscillatory modulations
of the bremsstrahlung and cyclotron luminosities.

This paper investigates the role of the lower boundary
in relation to the stability and emission properties
of post-shock flows.
This issue has not been thoroughly discussed in previous studies of
one-temperature (Saxton \etal 1998)
and two-temperature (Saxton \& Wu 1999, 2001) radiative shocks.
In an appendix, Imamura \etal (1996) stated that changing
the lower boudnary condition from constant-velocity to constant-pressure
did not affect their results with a power-law cooling function.
The present work explores
different analytic choices of the lower boundary condition,
in the presence of two cooling processes
and inequality between the electron and ion temperatures.
White dwarf accretion shocks 
are an illustrative case study.
The conditions considered herein include
constant pressure, density, flow-rate and temperature
at the white-dwarf surface,
or a boundary with nonzero longitudinal velocity
(\eg due to lateral ``leakage'' transverse to the magnetic field lines).

The plan of the paper is as follows.
Section~\ref{'section.analysis'} outlines the general formulation,
and section~\ref{'section.boundaryconditions'}
describes the cases of different boundary conditions investigated.
Section~\ref{'section.stability'} presents and discusses
the local and global stability properties.
Section~\ref{'section.luminosity'} examines
how the different lower boundary conditions
affect the modulation of the luminosities of the oscillating post-shock flow.
Section~\ref{'section.conclusions'} draws the conclusions.

\section{perturbative analysis}
\label{'section.analysis'}

The hydrodynamics of the accretion flow are governed by
the equations for
continuity of mass, momentum, total energy and electron energy:
\begin{equation}
\left({
{\partial\over{\partial t}} + \vb\cdot\nabla
}\right)\rho
+\rho(\nabla\cdot\vb) = 0.
\label{'eq.2d2t.continuity'}
\end{equation}
\begin{equation}
\rho\left({
{{\partial}\over{\partial{t}}}+\vb\cdot\nabla
}\right)\vb
=-\nabla{P}
\label{'eq.2d2t.momentum.xy'}
\end{equation}
\begin{equation}
\left({{{\partial}\over{\partial t}}+\vb\cdot\nabla}\right) P
-\gamma{{P}\over{\rho}}\left({{{\partial}\over{\partial t}}+\vb\cdot\nabla}\right)\rho
= -(\gamma -1)\Lambda
\label{'eq.2d2t.energy'}
\end{equation}
and
\begin{equation}
\left({{{\partial}\over{\partial t}}+\vb\cdot\nabla}\right) P\sube{}
-\gamma{{P\sube{}}\over{\rho}}\left({{{\partial}\over{\partial t}}+\vb\cdot\nabla}\right)\rho
= (\gamma-1)\left({\Gamma-\Lambda}\right)
\label{'eq.2d2t.electron'}
\end{equation}
where $\vb$, $\rho$, $P$ and $P\sube{}$
are the fluid velocity, density, total pressure and electron pressure,
and $\gamma$ is the adiabatic index (assumed equal to $5/3$ for an ideal monatomic gas).
The radiative loss term $\Lambda=\Lambda_{\rm br}+\Lambda_{\rm cy}$
comprises a sum of energy losses due to bremsstrahlung and cyclotron cooling.
The functional form of $\Lambda$ was described in Wu (1994).
\begin{equation}
\Lambda \equiv \Lambda_{_{\rm br}}+\Lambda_{_2}
=\Lambda_{\rm br}
\left[
1+\epsilon\subs{}\left({P\sube{} \over{P\sube{,s}}}\right)^\alpha
\left({\rho\over{\rho_{\rm s}}}\right)^{-\beta}
\right] \ , 
\label{'eq.lambda.total'}
\end{equation}
where $\rho\subs{}$ and $P\sube{,s}$
are the density and electron partial pressure at the shock,
and the bremsstrahlung cooling function is
$\Lambda_{\rm br}=A\rho^2\left({P\sube{}/\rho}\right)^{1/2}$
with
$A\approx3.9\times10^{16}$ in c.g.s.\ units (see \eg Rybicki \& Lightman 1979).
The parameter $\epsilon\subs{}$ is the efficiency of the second cooling process
compared to bremsstrahlung cooling,
evaluated at the shock.
For the case of cyclotron-emitting post-shock regions of accreting white dwarfs,
the indices are $\alpha=2.0$, $\beta=3.85$.

The term $\Gamma$ is the electron-ion energy exchange,
(\eg Spitzer 1962, Melrose 1986, Imamura \etal 1996).
\begin{equation}
\Gamma={{4\sqrt{2\pi}e^4n\sube{}n\subi{}\ln C}\over{m\sube{} c}}
\left[{
{\theta\subi{}-(m\sube{}/m\subi{})\theta\sube{}}
\over{(\theta\sube{}+\theta\subi{})^{3/2}}
}\right] \ , 
\end{equation}
where  $m_{\rm i,e}$,  $T_{\rm i,e}$, $\theta_{\rm i,e}=\kb T_{\rm i,e}/m_{\rm i,e}c^2$, 
and $n_{\rm i,e}$ are the ion and electron
masses, temperatures, dimensionless temperatures and number densities
of ions and electrons respectively. 
The other constants are $\kb$ the Boltzmann constant,
$c$ the speed of light,
$e$ the electron charge, 
and $\ln C$ is the Coulomb logarithm. 
 
Normalising the hydrodynamic quatities by their respective pre-shock values
($\vff$ the free-fall velocity,
$\rhoa$ the pre-shock density of the flow,
and $x\subs{0}$ the equilibrium shock height),
the variables of the stationary solution are defined as
$\tau_0=v/\vff$, $\pi_0=P/\rhoa\vff^2$ and $\pi\sube{}=P\sube{}/\rhoa\vff^2$
for the flow velocity, total pressure and electron pressure.
The hydrodynamic variables are functions of the vertical position coordinate
$\xi\equiv x/x\subs{}$, where $x\subs{}$ is the instantaneous shock height,
as distinct from the equilibrium value.
Then the cooling function and electron-ion exchange function are:
\begin{eqnarray}
\Lambda&=&(\rhoa\vff^3/x\subs{0})
\psic\tau_0^{3/2}{\pi\sube{}}^{1/2}
\left[{
1+\epsilon\subs{} f(\rhoa,P\sube{})
}\right]
\\
\Gamma&=&(\rhoa\vff^3/x\subs{0})
\psic\psiei\tau_0^{-5/2}{\pi\sube{}}^{-3/2}(1-\tau_0-2\pi\sube{})
\end{eqnarray}
where the constant $\psic=A x\subs{0}\rhoa/\vff$
and $\psiei$ parameterises the efficiency of electron-ion energy exchange
compared to the radiative cooling.
The conditions at the shock ($\xi=1$) are
$\tau_0=\frac14$, $\pi_0=\frac34$,
$\pi\sube{}={\frac38}(1+\sigma\subs{}^{-1})^{-1}$,
where $\sigma\subs{}\equiv (P\sube{}/P\subi{})\subs{}$
is the electron to ion pressure ratio at the shock.
The stationary state of the system is completely specified by the parameters
$(\sigma\subs{},\psiei,\epsilon\subs{})$.

As in Imamura \etal (1996) and Saxton \& Wu (1999),
the shock position is perturbed about its equilibrium position,
$x\subs{}=x\subs{0}+x\subs{1}e^{\omega t}$,
with complex frequency $\omega\equiv\delta\vff/x\subs{0}$.
The stationary solution
is obtained by separating the zeroth-order terms in the hydrodynamic equations,
involving the total pressure $\pi_0$
and electron pressure $\pi\sube{}$.

The perturbation of the post-shock structure
is described by a set of differential equations,
with the complex variables
$\lambda_\zeta(\xi)$, $\lambda_\tau(\xi)$,
$\lambda_\pi(\xi)$ and $\lambda\sube{}(\xi)$,
corresponding to perturbations of density, longitudinal velocity,
total pressure and electron pressure respectively.
The $\lambda$ functions are normalised by
$\varepsilon\equiv x\subs{1}\omega/\vff$,
the relative amplitude of the oscillation of the shock position.
The instantaneous density, velocity, total pressure and electron pressure
are 
\begin{eqnarray}
\rho(\xi,y,t)&=&\rhoa\cdot
\zeta_0(\xi) \left({1+\varepsilon\lambda_\zeta(\xi) e^{\omega t}
   }\right)\ ,
\\
v(\xi,y,t)&=&-\vff\cdot
\tau_0(\xi)
\left({1+\varepsilon\lambda_\tau(\xi) e^{\omega t}}\right),
\\
P(\xi,y,t)&=&\rhoa\vff^2\cdot
\pi_0(\xi) \left({1+\varepsilon\lambda_\pi(\xi) e^{\omega t}}\right) \ ,
\\
P\sube{}(\xi,y,t)&=&\rhoa\vff^2\cdot
\pi\sube{}(\xi) \left({1+\varepsilon\lambda\sube{}(\xi)
   e^{\omega t}}\right) \ .
\end{eqnarray}
Thus we obtain the first-order perturbed time-dependent equation
\begin{equation}
{d\over{d\tau_0}}
{
\left[
\begin{array}{c}
\lambda_\zeta\\
\lambda_\tau\\
\lambda_\pi\\
\lambda\sube{}
\end{array}
\right]
}
=
{1\over\tL}
{
\left[
\begin{array}{ccccc}
1&-1&0&{1/{\tau_0}}&0\\
{-{\gamma\pi_0}/{\tau_0}}&1&-{1/{\tau_0}}&0\\
\gamma&-\gamma&{1/{\pi_0}}&0\\
\gamma&-\gamma&{\gamma/{\tau_0}}&
 -{{(\gamma\pi_0-\tau_0)}/{\tau_0\pi\sube{}}}
\end{array}
\right]
}
{
\left[
\begin{array}{c}
F_1\\
F_2\\
F_4\\
F_5\\
\end{array}
\right]
}
\label{'eq.2d2t.matrix'}
\end{equation}
(see Saxton \& Wu 1999),
where $\tL\equiv(\gamma-1)\Lambda(x\subs{0}/\rhoa\vff^3)$
and $\tau_0\equiv-v_0/\vff$
are dimensionless versions of the cooling function
and accretion flow velocity.
The $F_1,\ldots,F_5$ terms are complex functions involving
the perturbed variables and the variables of the stationary solution.

\begin{figure}
\begin{center}
$
\begin{array}{cc}
\mbox{zero velocity}
&
\mbox{nonzero terminated velocity}
\\
\psfig{file=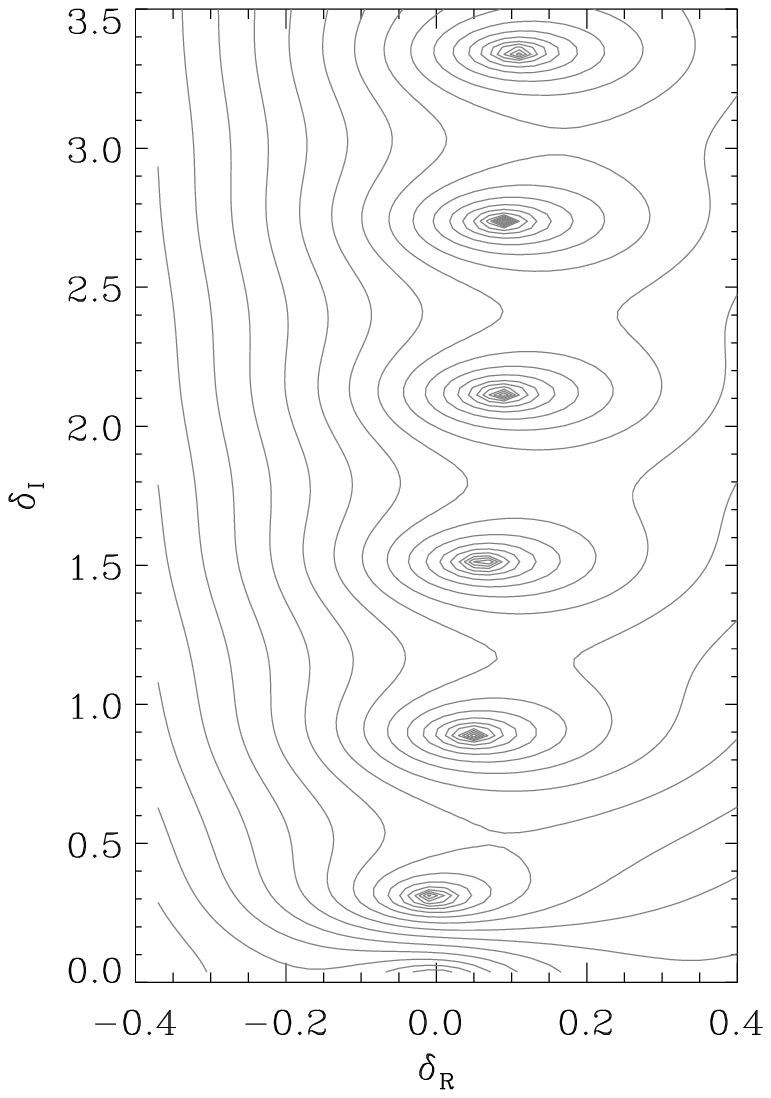,width=7cm}
&
\psfig{file=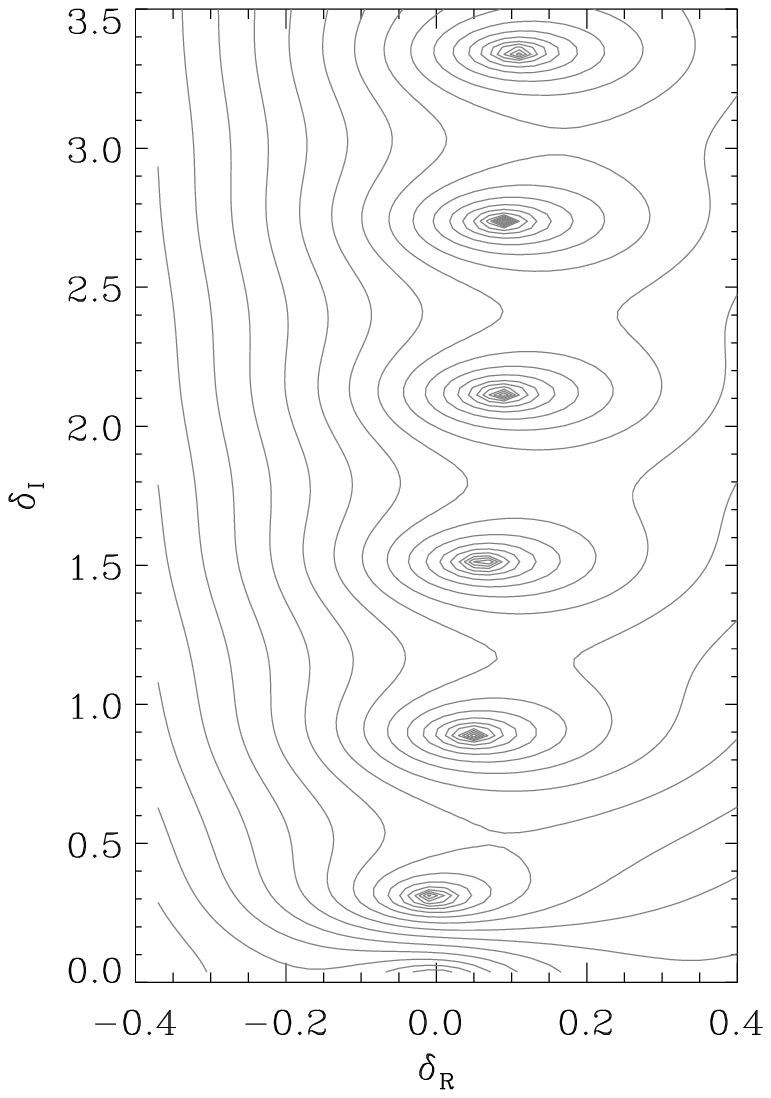,width=7cm}
\\
\psfig{file=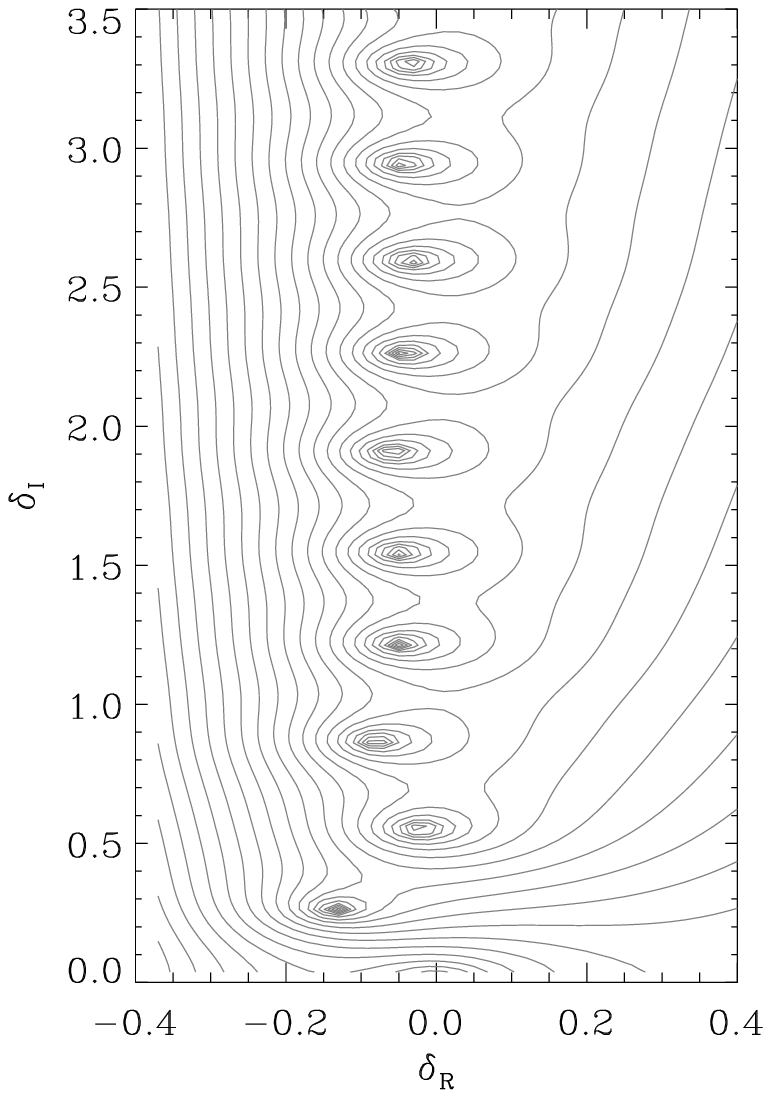,width=7cm}
&
\psfig{file=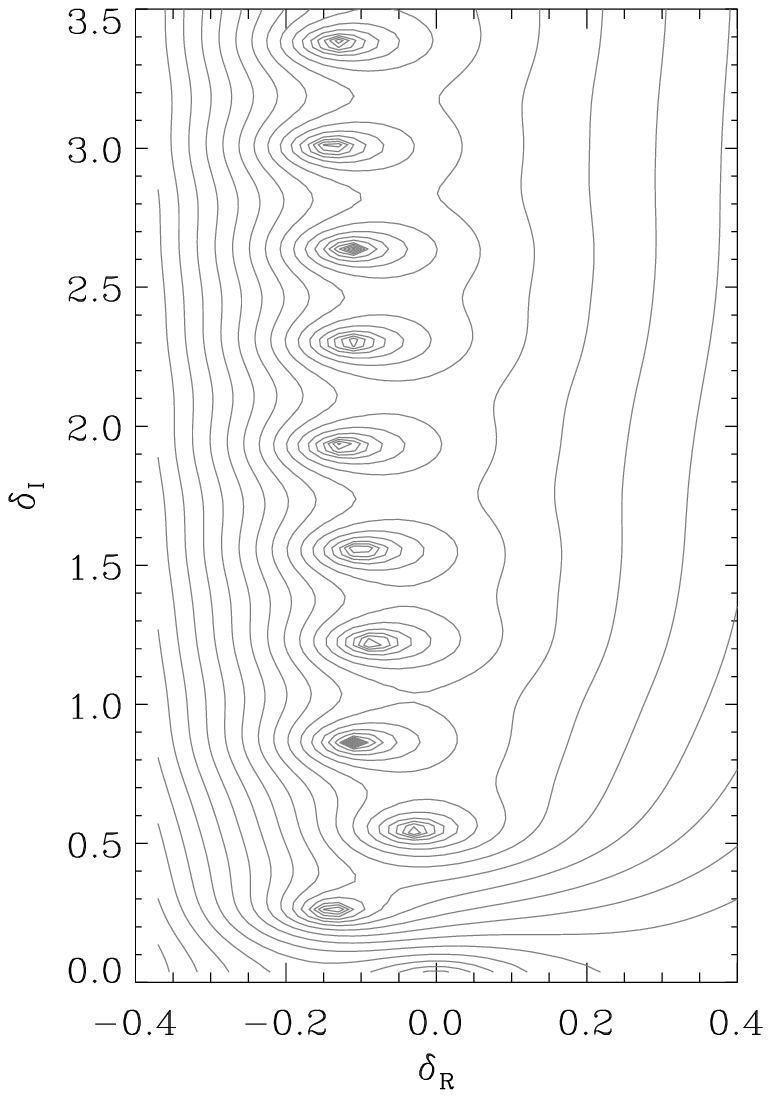,width=7cm}
\\
\end{array}
$
\caption{Complex frequency planes showing the eigenmodes
of systems with parameters
$(\sigma\subs{},\psiei,\epsilon\subs{})=(0.5,0.5,0)$ in the top row
and
$(\sigma\subs{},\psiei,\epsilon\subs{})=(0.5,0.5,100)$ in the bottom row.
The lower boundary condition is
$\lambda_\tau=0$, $\tau_0=0$ in the left column,
and $\lambda_\tau=0$, $\tau_0=0.01$ in the right column.
The contour levels are at logarithmic intervals
of the function $1/|\lambda_\tau|$;
the modes are at the peaks.
The real part of the an eigenvalue, $\dR$,
describes the instability of a mode:
positive indicates instability.
The imaginary part of the eigenvalue, $\dI$,
is proportional to the frequency of oscillation.
}
\label{'fig.eigenplanes'}
\end{center}
\end{figure}

The differential equations are integrated
from the shock down to the lower boundary.
Values of $\delta$ which satisfy the appropriate boundary conditions
are the eigenfrequencies.
The complex eigenvalues $\delta=\dR+i\dI$
describe the oscillatory properties of the shock
in different modes.
The real part, $\dR$, is a growth term:
if $\dR>0$ then the shock is unstable to oscillations
in the corresponding mode.
The imaginary part, $\dI$,
is proportional to the oscillatory frequency of the mode.
Figure~\ref{'fig.eigenplanes'} shows examples of how
the eigenplanes of the perturbed variables
$-\log|\lambda_\tau|$
evaluated for a given set of boundary conditions.
The $\lambda$-profiles corresponding to the $\delta$ eigenvalues
are the eigenfunctions of the post-shock structure's local oscillatory response
to the shock oscillations.

For each cooling process, the luminosity response is calculated
by multiplying the stationary-state cooling function
and corresponding perturbation,
and then integrating their product over the entire post-shock region.
The luminosity response of a mode,
is expressed in terms of the stationary-solution luminosity
and normalised by the size of the shock-height oscillation, \ie
\begin{eqnarray}
{{ L_{\rm br,1} }\over{ \varepsilon L_{\rm br,0} }}&=&
{{ \int_0^1\Lambda_{\rm br}\lambda_{\rm br}d\xi }
\over{ \int_0^1\Lambda_{\rm br} d\xi }} \ ,
\nonumber \\
{{ L_{\rm cy,1} }\over{ \varepsilon L_{\rm cy,0} }}&=&
{{ \int_0^1\Lambda_{\rm cy}\lambda_{\rm cy}d\xi }
\over{ \int_0^1\Lambda_{\rm cy} d\xi }}
\label{'eq.def.luminosity'}
\end{eqnarray}
where $\lambda_{\rm br}={\frac32}\lambda_\zeta+{\frac12}\lambda\sube{}$
and $\lambda_{\rm cy}=-2.35\lambda_\zeta+2.5\lambda\sube{}$
(see Saxton \& Wu 2001).
The absolute value of the ratio (\ref{'eq.def.luminosity'})
is the relative amplitude of
the bremsstrahlung(cyclotron) luminosity oscillation,
and the phase $\Phi_{\rm br}$ (or $\Phi_{\rm cy}$)
is relative to the oscillatory phase of the shock motion.

\section{Boundary conditions}
\label{'section.boundaryconditions'}

\subsection{shock jump condition}

The boundary conditions at the shock ($\xi=1$) are determined by
the strong-shock Rankine-Hugoniot conditions.
It can be shown that the stationary-solution conditions are
$\tau_0=\frac14$, $\pi_0=\frac34$.
If the ratio of electron and ion pressures is parameterised as $\sigma\subs{}$
then the electron pressure condition is
$\pi\sube{}={\frac34}/(1+\sigma\subs{}^{-1})$.
The conditions on the perturbed variables
are derived in the frame that is comoving with the shock
(Imamura \etal 1996, see also appendices of Saxton \etal 1997),
$\lambda_\zeta=0$, $\lambda_\tau=-3$,
and $\lambda_\pi=\lambda_\sube{}=2$.
The system parameters are the same as defined in Saxton \& Wu (1999):
$\epsilon\subs{}$ is the efficiency of cyclotron cooling
relative to bremsstrahlung cooling at the shock;
$\sigma\subs{}$ is the ratio of electron to ion pressures at the shock;
$\psiei$ specifies the efficiency of electron-ion energy exchange
relative to the total efficiency of radiative cooling.
The shock boundary conditions are the same in all cases studied;
alternative conditions at the lower boundary are considered.

\subsection{lower boundary conditions}

The post-shock flow has a lower boundary, labelled $\xi=0$,
and the form and value of the boundary conditions there
depend on the nature of the system under consideration.
For example,
the boundary of cooled gas far downstream from an interstellar medium shock
and the boundary formed by a stellar surface in an accretion flow
may impose different conditions on the hydrodynamic variables.

\subsubsection{``perfect'' stationary-wall}
\label{'bc.perfect'}

If the flow velocity vanishes at the lower boundary
then the condition is
$\tau_0=0$ at $\xi=0$.
This is sometimes known as the ``stationary-wall'' condition.
The additional choice of $\lambda_\tau=0$
means that the flow velocity is absolutely zero
and does not oscillate at the lower boundary.
The ``perfect stationary-wall'' boundary condition
is the conventional choice in many stability analyses of radiative shocks
(\eg Chevalier \& Imamura 1982, T\'{o}th \& Draine 1993, Imamura \etal 1996,
Saxton \etal 1998).
Bertschinger (1986) demonstrated that this condition also applies
to an ISM radiative shock, at the inner boundary which is defined
as the place where gas is cooled to a point where total cooling is zero.

Wolff,~Gardner~\&~Wood~(1989)
carried out numerical studies of accreetion onto white dwarfs
where cyclotron cooling does not dominate, 
and bremsstrahlung cooling,
Compton cooling, electron-ion energy exchanges 
and electron thermal conduction are included.
They considered a ``perfect stationary-wall'' condition 
at the lower boundary,
with additional constraints:
(i) a constant nonzero electron temperature
$T_e=T_{\rm cut}$, the cut-off value where the total cooling $\Lambda=0$;
(ii) a zero gradient in temperature,
$\partial T\sube{}/\partial r = 0$.
Thes presence of conduction 
and the alternative form of $\Lambda$
must provide a solution for time-independent flow structure
that is significantly different from that considered in the present work,
which focusses on the role of lower boundary conditions
on the perturbed variables.
The effects of boundary conditions on stability properties
of structures where conduction and finite cooling cut-offs 
are physically important
comprises a significant topic for future investigation.

\subsubsection{modified stationary-wall}
\label{'ss.modified'}

The $\lambda_\tau=0$ fixed-velocity assumption is not the only choice
of lower boundary condition which is consistent with
the stationary solution that provides $\tau_0=0$ at $\xi=0$.
There are alternative physical propositions.

\label{'bc.pressure'}
The stationary-solution pressure at the lower boundary
is a well-defined value, $\pi_0=1$.
The condition $\lambda_\pi=0$
means that the total pressure of the flow does not oscillate
at the lower boundary.
This is a kind of hydrostatic condition.
In the case of accreting white dwarfs,
this condition would mean that the pressure wave of the shock oscillation
does not enter the white dwarf atmosphere.

\label{'bc.density'}
The condition $\lambda_\zeta=0$
means that the density does not oscillate at the lower boundary.
\label{'bc.flowrate'}
The condition $\lambda_\zeta+\lambda_\tau=0$
stands for constancy of the mass flow rate $\rho v$
at the lower boundary.
The physical interpretation of these two choices has a subtle complication:
in the present treatment the density grows indefinitely as $\xi\rightarrow 0$.
However this problem would be alleviated
in a modified geometry
(\eg taking account of lateral flow from the sides of the accretion column)
which will be investigated in future studies.
(The present model applies to white dwarf accretion shocks so long as
lateral spreading of accreted gas is much slower than infall at the shock,
and buildup of material at the base is slow 
on the cooling and oscillatory timescales.)

\label{'bc.temperature'}
Other physical lower boundary conditions
are related to the thermal structure of the post-shock flow.
The electron and ion temperatures must be equal at the white dwarf surface,
which implies that $\lambda\sube{}-\lambda_\pi=0$
is a necessary boundary condition.
However this condition is satisfied for every choice of $\delta$,
since the functional form of the electron-ion energy exchange $\Gamma$
ultimately exceeds the bremsstrahlung cooling $\Lambda_{\rm br}$
in the high-density, low-temperature region near the lower boundary
(see \eg Saxton \& Wu 2001).
Another choice would be to fix the flow temperature
at the lower boundary,
assuming that the gas at the base of the flow
is in thermal equilibrium with the white dwarf,
and that the white dwarf surface is a perfect heat-bath/heat-sink.
Constancy of $T\propto P/\rho$ implies a condition of
$\lambda_\pi-\lambda_\zeta=0$.
The lower-boundary temperature is zero in the stationary solution
for the present example,
so this choice of condition on the perturbed variables
describes the lower boundary as being a perfect node in
the oscillation of the temperature profile.

\subsubsection{nonzero terminated velocity}
\label{'bc.truncated'}

This paper also considers an example in which
the lower boundary has a non-zero flow velocity relative to the shock.
This condition describes a system where
cooling at the base does not compress the gas indefinitely,
and the the accretion shock moves outwards
from the white dwarf surface
(until non-linear effects become important
or the shock becomes indistinct).
Alternatively the boundary condition could describe
a surface at which post-shock material is removed from the system
at a nonzero rate,
like flow into a ``leaky'' bucket.
The illustrative choice considered here is
$\tau_0\rightarrow 0.01$ rather than zero,
and $\lambda_\tau=0$ at the lower boundary.
In other words,
the shock propagates away from the white dwarf surface
at a velocity that is $0.01$ times the free-fall velocity of pre-shock matter,
or else gas is removed from the accretion stream where it contacts the stellar surface
at incident velocity $0.01\vff$.

\begin{table}
\renewcommand{\baselinestretch}{1.0}
\caption{
The boundary conditions studied in this paper,
in terms of the stationary and perturbed variables.
}
\begin{center}
$
\begin{array}{rlll}
\mbox{case}&\mbox{stationary variables}&\mbox{perturbed variables}&\mbox{note}
\\
\hline\\
1&\tau_0=0,\hskip 15pt \pi_0=1	&\lambda_\tau=0
&\mbox{\S\ref{'bc.perfect'} ``perfect stationary-wall''}
\\
2&\tau_0=0,\hskip 15pt \pi_0=1	&\lambda_\pi=0
&\mbox{\S\ref{'bc.pressure'} fixed pressure}
\\
3&\tau_0=0,\hskip 15pt \pi_0=1	&\lambda_\zeta+\lambda_\tau=0
&\mbox{\S\ref{'bc.flowrate'} fixed flow rate}
\\
4&\tau_0=0,\hskip 15pt \pi_0=1	&\lambda_\zeta=0
&\mbox{\S\ref{'bc.density'} fixed density}
\\
5&\tau_0=0,\hskip 15pt \pi_0=1	&\lambda_\pi-\lambda_\zeta=0
&\mbox{\S\ref{'bc.temperature'} fixed temperature}
\\
6&\tau_0=0.01, \pi_0=0.99	&\lambda_\tau=0
&\mbox{\S\ref{'bc.truncated'} nonzero terminated velocity;}
\\
&&&\mbox{\hskip 30pt ``leaky flow''}
\\
\hline
\end{array}
$
\end{center}
\end{table}

\section{Stability properties}
\label{'section.stability'}

\subsection{eigenfunctions in the standard case}

In the ``perfect stationary wall'' case,
there exist amplitude nodes and antinodes
in the eigenfunctions for perturbed density ($|\lambda_\zeta|$),
and also for electron pressure ($|\lambda\sube{}|$)
in cases when the two-temperature effects are important.
The number of nodes depends on the harmonic number of the mode.
Phases of the eigenfunctions change abruptly at the node locations,
and wind gradually in intervening regions.

Apart from the perturbed velocity 
($\lambda_\tau=0$ by the boundary condition)
the perturbed variables are not explicitly determined at the lower boundary,
but have finite amplitudes within an order of magnitude of
their typical amplitudes throughout the whole post-shock region.
The perturbed electron pressure and total pressure variables
always meet at the same value at the lower boundary,
regardless of the boundary condition,
because the electron and ion temperatures equilibrate at the lower boundary.
For comprehensive discussion of the eigenfunctions,
see Saxton (1999) and Saxton \& Wu (2000).

\subsection{sensitivity of eigenfunctions}

Saxton (1999) \S5.4.1-5.4.2 considered the relationships
between eigenvalues and local features of the corresponding
eigenfunctions.
The relevant findings were as follows.

\label{'eigenfn.small.offset'}
\label{'eigenfn.at.bc'}
For arbitrary alterations of the lower boundary condition
resulting in a small offset of the eigenvalue,
$|\Delta\delta|=0.01$,
the eigenfunctions alter significantly in the region $\xi\LS 0.05$
near the lower boundary,
but remain essentially unchanged throughout the majority of
the post-shock region ($0.05\LS\xi\leq 1$).
In the sensitive base region,
offsets from the ``perfect stationary wall'' condition
yield steeper gradients in the amplitudes and phases of all
the $\lambda$-variables.
The amplitudes $|\lambda_\zeta|$ and $|\lambda_\tau|$ 
may differ their ordinary values by over an order of magnitude;
the pressure amplitudes $|\lambda_\pi|$ and $|\lambda\sube{}$
differ by factors of a few.

At more extremely different boundary conditions,
resulting in eigenvalues that are intermediate between modes
of the ``perfect stationary-wall'' case,
the sensitive, base region ($\xi\LS0.03$) 
with steep gradients in the
$\lambda$-functions persists.
Amplitudes are enhanced in regions up to $\xi\LS0.85$,
but this is not as significant as the efffects in the base region.
The number of antinodes and nodes resmbleare comparable to those
of the next higher modes of the ``perfect stationary-wall'' case,
but their positions are completely different.

Thus small offsets of the global frequency and stability properties
alter the local behaviour of the lowest few percent
of the post-shock flow,
but have no significant effects in regions of the flow further 
from the lower boundary.

\subsection{eigenvalues: stability and frequencies}

The previous subsection reviewed
how eigenfunctions of the hydrodynamic variables
generally respond to alterations of lower boundary conditions
resulting in small and large alterations 
of the frequencies and stability properties. 
This section details the effects on the eigenvalues,
the mode frequencies and stabilitiy properties,
of specific alternatives for the lower boundary condition.

\subsubsection{``perfect stationary-wall''}

The general stability and frequency results
for the standard ``perfect stationary-wall'' lower boundary condition
are as follows
(see \eg Chevalier \& Imamura 1982, Imamura \etal 1996,
Saxton \etal 1997, 1998 and Saxton \& Wu 1999, 2000).
When bremsstrahlung emission is the only cooling process
the fundamental mode is stable,
and the overtones are unstable,
with instability $\dR$ tending to increase with harmonic number.
When two-temperature effects are unimportant ($\psiei>1$),
increasing the cyclotron cooling efficiency ($\epsilon\subs{}$)
stabilises each mode.
Some modes stabilise more readily than others
and the trend of $\dR$ increasing with harmonic number breaks down
more extensively as cyclotron cooling becomes more efficient.
In the absence of two-temperature effects,
the oscillatory frequencies are quantised like those of 
a pipe open at one end
(T\'{o}th \& Draine 1993;
Saxton \etal 1997, 1999). 
Increasing $\epsilon\subs{}$ reduces
the frequency intervals of between consecutive modes
(Saxton \etal 1997; Saxton \& Wu 1999).
When two-temperature effects are important ($\psiei\LS1$),
greater $\epsilon\subs{}$ may actually destabilise modes
(Imamura \etal 1996)
and the frequency sequence becomes more like that of a doubly-open
or doubly-closed pipe
(Saxton \& Wu 1999).

Small offsets of the $\delta$ values
may produce large changes in the $\lambda$-profiles
in the region near the lower boundary.
The affected region is only a few percent of the entire post-shock flow.
For larger $\delta$ offsets the $\lambda$-profiles
are qualitatively similar to
the eigenfunctions of the nearest higher mode, except that:
(1) the perturbed variables change abruptly
near the lower boundary, as for small offsets;
(2) throughout most of the post-shock region
there is a general increase of amplitude with decreasing $\xi$;
and (3) the nodes and antinodes are located at different positions.
Thus systems with eigenfrequencies and stability properties
very different from the stationary-wall case
can nonetheless have similar local oscillatory responses
throughout most of the post-shock region except for a small base region.

\subsubsection{modified stationary wall}

Table~\ref{'table.eigenvalues'}
presents eigenvalues
for the alternative boundary conditions described in
\S\ref{'ss.modified'}: 
constant pressure, flow rate, density or temperature.
Compared to the ``perfect stationary-wall'' case,
all of the modified stationary-wall conditions
yield complex $\delta$ values that differ by only a few percent.
The changes of $\delta$ are smaller than or comparable to
the idealised ``small offset'' cases
(\S\ref{'eigenfn.small.offset'} above;
also Saxton 1999, \S5.4.1)).

Thus the alteration of the lower boundary condition on perturbed variables
has no significant effect on the global stability properties
and oscillatory frequencies of the modes.
The nature and efficiencies of the radiative cooling processes 
and the electron-ion energy exchange
have a greater influence on the frequencies and stability properties
of the oscillating shocks than the lower boundary conditions.
Observed frequency ratios and damping characteristics
of shock oscillations (\eg in accretion onto magnetic white dwarfs)
is expected to be 
practically independent of the lower boundary condition.
(However the lower boundary may still have significant effects upon the
modulation of bremsstrahlung and cyclotron luminosities,
as discussed in \S5.)

\subsubsection{nonzero terminated velocity}

If the flow velocity terminates with a nonzero value, $\tau_0=0.01$,
at the lower boundary (\S\ref{'bc.truncated'}),
the oscillatory frequencies are essentially unchanged,
and the stability properties change slightly.
For the cases studied,
when bremsstrahlung emission is the only cooling process
the difference between zero-velocity termination
and nonzero-velocity termination
is a few times $0.001$ in both $\dR$ and $\dI$;
for $\epsilon\subs{}=100$
the changes are a few times $0.01$.
Changing from zero-velocity termination to nonzero-velocity termination
at the lower boundary
made every mode more stable.
Higher-order modes are more significantly affected,
and the effects are greatest when $\epsilon\subs{}$ is higher.
As seen in Figure~\ref{'fig.eigenplanes'},
the sequence of the modes' stability properties is qualitatively different
for the zero- and nonzero-terminated velocity boundary conditions
in the cyclotron-dominated extreme
$(\sigma\subs{},\psiei,\epsilon\subs{})=(0.5,0.5,100)$.

The increased stability of modes
in the cases of cyclotron-dominated cooling
may occur because the conditions at the lower boundary
are not as dense as in the ``perfect stationary-wall'' case.
Since the highest densities near the lower boundary are less extreme,
the bremsstrahlung cooling dominates over a smaller region
at the base of the post-shock flow.
As bremsstrahloung radiative cooling is responsible for 
the thermal instability that drives the shock oscillations,
it is qualitatively sensible that the ``nonzero terminated velocity'' case
has more stable modes than the ``perfect stationary-wall'' case.

\begin{table}
\renewcommand{\baselinestretch}{1.0}
\caption{
Eigenvalues for four alternative choices of the lower boundary condition
in the perturbed variables,
but with the standard conditions on the stationary solution
($\tau_0=0$, $\pi_0=1$, $\pi\sube{}={\frac12}$ at $\xi=0$).
Values shown are for the first six harmonics, $n=1\ldots6$.
}
{\tiny
\begin{center}
$
\begin{array}{ccccccc}
& & &\lambda_\pi=0&\lambda_\zeta+\lambda_\tau=0&\lambda_\zeta=0&\lambda_\pi-\lambda_\zeta=0\\
\\
\sigma\subs{}&\psiei&\epsilon\subs{}&
\begin{array}{cc}\hline\\ \dR&\dI\\~~~~~~~~~~&~~~~~~~~~~\end{array}
&
\begin{array}{cc}\hline\\ \dR&\dI\\~~~~~~~~~~&~~~~~~~~~~\end{array}
&
\begin{array}{cc}\hline\\ \dR&\dI\\~~~~~~~~~~&~~~~~~~~~~\end{array}
&
\begin{array}{cc}\hline\\ \dR&\dI\\~~~~~~~~~~&~~~~~~~~~~\end{array}
\\
\hline
\\
\begin{array}{r}
0.5\\ \\ \\ \\ \\ \\
\end{array}&
\begin{array}{r}
0.5\\ \\ \\ \\ \\ \\
\end{array}&
\begin{array}{r}
0\\ \\ \\ \\ \\ \\
\end{array}&
\begin{array}{cc}
	\begin{array}{r}
	-0.006\\0.048\\0.063\\0.087\\0.091\\0.109
	\end{array}&
	\begin{array}{r}
	0.305\\0.890\\1.507\\2.111\\2.728\\3.334
	\end{array}
\end{array}&%
\begin{array}{cc}
	\begin{array}{r}
	-0.007\\0.047\\0.060\\0.084\\0.086\\0.104
	\end{array}&
	\begin{array}{r}
	0.307\\0.895\\1.514\\2.121\\2.741\\3.351
	\end{array}
\end{array}&%
\begin{array}{cc}
	\begin{array}{r}
	-0.006\\0.048\\0.061\\0.086\\0.088\\0.107
	\end{array}&
	\begin{array}{r}
	0.306\\0.892\\1.511\\2.117\\2.736\\3.345
	\end{array}
\end{array}&%
\begin{array}{cc}
	\begin{array}{r}
	-0.006\\0.048\\0.061\\0.086\\0.088\\0.107
	\end{array}&
	\begin{array}{r}
	0.306\\0.892\\1.511\\2.117\\2.736\\3.345
	\end{array}
\end{array}
\\
\\
\begin{array}{r}
0.5\\ \\ \\ \\ \\ \\
\end{array}&
\begin{array}{r}
0.5\\ \\ \\ \\ \\ \\
\end{array}&
\begin{array}{r}
1\\ \\ \\ \\ \\ \\
\end{array}&
\begin{array}{cc}
	\begin{array}{r}
	-0.064\\-0.018\\-0.013\\0.013\\0.011\\0.035
	\end{array}&
	\begin{array}{r}
	0.331\\0.820\\1.385\\1.921\\2.481\\3.030
	\end{array}
\end{array}&%
\begin{array}{cc}
	\begin{array}{r}
	-0.066\\-0.020\\-0.017\\0.009\\0.003\\0.028
	\end{array}&
	\begin{array}{r}
	0.333\\0.825\\1.393\\1.935\\2.496\\3.054
	\end{array}
\end{array}&%
\begin{array}{cc}
	\begin{array}{r}
	-0.065\\-0.019\\-0.016\\0.011\\0.006\\0.031
	\end{array}&
	\begin{array}{r}
	0.332\\0.823\\1.389\\1.930\\2.490\\3.047
	\end{array}
\end{array}&%
\begin{array}{cc}
	\begin{array}{r}
	-0.065\\-0.019\\-0.016\\0.011\\0.006\\0.031
	\end{array}&
	\begin{array}{r}
	0.332\\0.823\\1.389\\1.930\\2.490\\3.047
	\end{array}
\end{array}
\\
\\
\begin{array}{r}
0.5\\ \\ \\ \\ \\ \\
\end{array}&
\begin{array}{r}
0.5\\ \\ \\ \\ \\ \\
\end{array}&
\begin{array}{r}
100\\ \\ \\ \\ \\ \\
\end{array}&
\begin{array}{cc}
	\begin{array}{r}
	-0.128\\-0.023\\-0.070\\-0.045\\-0.072\\-0.063
	\end{array}&
	\begin{array}{r}
	0.269\\0.526\\0.859\\1.164\\1.505\\1.859
	\end{array}
\end{array}&%
\begin{array}{cc}
	\begin{array}{r}
	-0.137\\-0.024\\-0.085\\-0.053\\-0.056\\-0.065
	\end{array}&
	\begin{array}{r}
	0.269\\0.554\\0.876\\1.226\\1.556\\1.919
	\end{array}
\end{array}&%
\begin{array}{cc}
	\begin{array}{r}
	-0.135\\-0.022\\-0.081\\-0.049\\-0.051\\-0.059
	\end{array}&
	\begin{array}{r}
	0.264\\0.551\\0.868\\1.218\\1.547\\1.907
	\end{array}
\end{array}&%
\begin{array}{cc}
	\begin{array}{r}
	-0.135\\-0.022\\-0.081\\-0.049\\-0.051\\-0.059
	\end{array}&
	\begin{array}{r}
	0.264\\0.551\\0.868\\1.218\\1.547\\1.907
	\end{array}
\end{array}
\\
\\
\hline
\end{array}
$
\end{center}
}
\label{'table.eigenvalues'}
\end{table}

\section{Luminosity response}
\label{'section.luminosity'}

Oscillations of the post-shock flow structure
modulate the emitted radiation.
Rapid $\sim1{\rm Hz}$ shock oscillations 
were observataionally inferred in 
accretion shocks of white dwarfs in some AM Herculis (AM~Her) systems
by optical-infrared observations of the oscillating cyclotron luminosities
(Middleditch 1982; 
Imamura \& Steiman-Cameron 1986;
Larsson 1987, 1989;
Ramseyer \etal 1993;
Middleditch, Imamura \& Steiman-Cameron 1997).
However the shock oscillations of these systems
may not produce comparable oscillations of bremsstrahlung luminosity.
For some systems the X-ray luminosity responses 
are subject to upper limits of a few percent
(Wolff \etal 1999;
Beardmore \& Osborne 1997).
It is therefore interesting to predict
the oscillatory responses of the bremsstrahlung and cyclotron 
total luminosities implied by the analytically obtained modes,
and their phasing properties 
(Saxton \& Wu 2001).

Wolff \etal (1991) and Wood \etal (1992)
carried out numerical investigations of noise-drive QPOs
in cases where the magnetic field strength is low enough
that the cyclotron radiative cooling does not dominate.
They found that modes that are suppressed in linear analyses
may be sustained by accretion noise
and show significant luminosity responses.
The luminosity responses were found to be linear,
and bremsstrahlung and cyclotron amplitudes were comparable
although bremsstrahlung signatures of the modes are broader
and therefore less distinct in the frequency power spectrum.

The present work extends the previous analysis
to consider how the luminosity oscillations
may depend on the character of the lower boundary.
The relative suppression ofrapid oscillations in the X-rays
may probe the nature of the boundary
at the bottom of the post-shock accretion flow onto the white dwarf 
in AM~Her systems.

The different radiative cooling processes
and the electron-ion energy exchange are strong
in different regions of the post-shock flow.
Bremsstrahlung cooling depends on a high power of density,
and therefore it peaks in the cold, high-density region 
near the lower boundary.
Cyclotron cooling depends on a high power of temperature,
and therefore most of the cyclotron emission
comes from the hot region near the shock.
Therefore the local oscillations of hydrodynamic variables
in the region near the lower boundary are
effective at modulating the bremsstrahlung luminosity,
whereas the oscillatory behaviour of the flow near the shock
is most effective at modulating the cyclotron luminosity.
Thus the bremsstrahlung luminosity and the cyclotron luminosity
have different responses to the shock oscillation of any given mode.
Furthermore,
because different modes have different profiles of 
nodes, antinodes and phases between the upper and lower boundaries,
the oscillatory responses of the bremsstrahlung and cyclotron luminosities
must differ between modes.

As shown in Saxton (1999),
the oscillatory eigenfunctions
are sensitive to changes of the lower boundary condition
in the region close to the boundary ($\xi\LS0.03$)
but insensitive in the rest of the post-shock flow
($0.03\LS\xi\le 1$).
Since bremsstrahlung emission peaks near the lower boundary
and cyclotron emission peaks near the shock,
it is to be expected that the integrated bremsstrahlung luminosity
is sensitive to alterations of the lower boundary condition,
but the cyclotron luminosity is not.
The bremsstrahlung luminosity response
is additionally sensitive to the lower boundary condition
by way of the large amplitudes in the density eigenfunction $\lambda_\zeta$
that can occur at low $\xi$ for some $\delta$ frequency values.
The calculated luminosity responses shown in
Table~\ref{'table.luminosity.altbc'}
confirm this prediction
when the ``perfect stationary-wall'' condition
($\lambda_\tau=0$) at the lower boundary
is replaced by the alternative conditions
described in \S\ref{'ss.modified'} and \S\ref{'bc.truncated'}.
The bremsstrahlung luminosity amplitudes and phasings change significantly
under alterations of the lower boundary condition,
but the cyclotron luminosity response is essentially unaffected.

Because of the insensitivity of cyclotron luminosity responses to
the lower boundary condition,
the following discussion concentrates on the bremsstrahlung response alone.

\subsection{``perfect stationary-wall''}

As discussed for the choice of a ``perfect stationary-wall'' lower boundary
in Saxton \& Wu (1999),
the modes' luminosity responses were found to be insensitive to
$(\sigma\subs{},\psiei)$
when bremsstrahlung cooling dominates ($\epsilon\subs{}=0$).
However there are no obvious relationships between the luminosity responses
and stability properties of the modes
for diverse cases of the system parameters.
It was found that the cyclotron luminosity oscillation
tends to lag the bremsstrahlung luminosity oscillation
by approximately $0.6\pi$ radians for the fundamental mode,
but the overtones lack obvious relationships between their bremsstrahlung
and cyclotron responses.

\begin{figure}
\begin{center}
\psfig{file=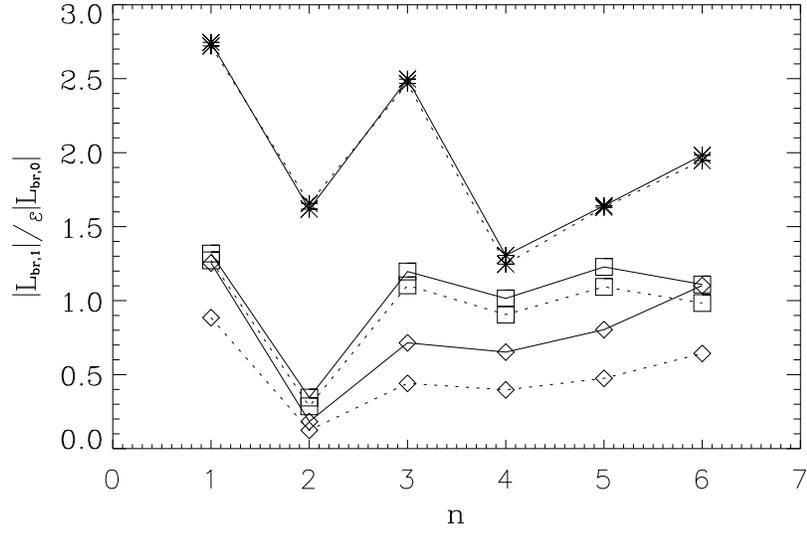,height=8cm}
\end{center}
\caption{
Comparison of bremstrahlung luminosity oscillatory amplitude
for cases with ``perfect stationary-wall'' ($\lambda_\tau=0$, solid lines)
and fixed-density ($\lambda_\zeta=0$, dotted lines) lower boundary conditions.
Values for the first six harmonics are marked with
stars, squares and diamonds
for $\epsilon\subs{}=0,1,100$ respectively.
The other system parameters are $(\sigma\subs{},\psiei)=(0.5,0.5)$.
For given system parameters,
the $|L_{\rm br,1}|/\varepsilon L_{\rm br,0}$ values
in the fixed-density case are approximately proportional to
corresponding values in the ``perfect stationary-wall'' case.
}
\label{'fig.zetaVtau'}
\end{figure}

\subsection{fixed-density or fixed-temperature}

The bremssttrahlung luminosity responses
are identical to within a few percent
for the cases of
fixed-density ($\lambda_\zeta=0$)
and fixed-temperature $\lambda_\pi-\lambda_\zeta=0$)
boundary conditions.
Thus in this sense an absence of oscillation in density at the lower boundary
is almost equivalent to an absence of oscillation in temperature.
For each choice of $(\sigma\subs{},\psiei,\epsilon\subs{})$
there is an approximately constant ratio between
the $|L_{\rm br,1}|/\varepsilon L_{\rm br,0}$ values 
in the fixed-density and fixed-temperature cases
as compared to corresponding modes in the ``perfect stationary wall'' case
(see Figure~\ref{'fig.zetaVtau'}).
For $(\sigma\subs{},\psiei,\epsilon\subs{})=(0.5,0.5)$
the ``perfect stationary-wall''
and fixed-density/temperature $|L_{\rm br,1}|$ values
are in the approximate ratio of $1.0$ for $\epsilon\subs{}=0$,
$1.0-1.2$ for $\epsilon\subs{}=1$,
and $1.3-1.7$ for $\epsilon\subs{}=100$.
The phases of the bremsstrahlung luminosity responses
are approximately equal for each mode
in the fixed-density/fixed-temperature cases
and the ``perfect stationary-wall'' case.
For the studied modes
where $(\sigma\subs{},\psiei)=(0.5,0.5)$,
the phase differences are
$\LS0.02\pi$ radians for $\epsilon\subs{}=0$,
$\LS0.03\pi$ radians for $\epsilon\subs{}=1$,
and
$\LS0.04\pi$ radians for $\epsilon\subs{}=100$,

\subsection{fixed-pressure}

With the present choices of system parameters,
the bremsstrahlung luminosity of each mode
is greater under the fixed-pressure condition ($\lambda_\pi=0$)
than under the ``perfect stationary-wall'' condition ($\lambda_\tau=0$).
Modes of higher harmonic number
tend to have higher amplitudes of bremsstrahlung luminosity response.
(One exception is the $n=3$ mode when
$(\sigma\subs{},\psiei,\epsilon\subs{})=(0.5,0.5,100)$,
which is also unusual because the instability eigenvalue $\dR$
departs from the usual trend of $\dR$ increasing with harmonic number $n$.)
The fixed-pressure and ``perfect stationary-wall'' lower boundary conditions
also produce different bremsstrahlung luminosity phases.
As shown in the top-left and bottom-left blocks
of Table~\ref{'table.luminosity.altbc'},
the bremsstrahlung luminosity phases for the first six harmonics differ by
$\sim0.3\pi - 0.8\pi$ for
$(\sigma\subs{},\psiei)=(0.5,0.5)$ and $\epsilon\subs{}=1,100$.
The change of boundary condition makes less difference 
($\LS 0.3\pi$ in phase) for the same modes
when cyclotron cooling is absent ($\epsilon\subs{}=0$).

\subsection{fixed flow-rate}

In the cases with fixed flow-rate boundary condition
($\lambda_\zeta+\lambda_\tau=0$)
the bremsstrahlung luminosity response amplitudes
generally decrease with harmonic number, $n$.
The exceptions, for the choices of system parameters studied here,
are the modes which depart from the trend of
$\dR$ increasing with harmonic number, $n$.
There is no clear relationship between the phases
of bremsstrahlung luminosity oscillation
in the ``perfect stationary-wall'' and fixed-flow cases,
however the phase of each mode in the fixed-flow case
is approximately in antiphase to that in the corresponding fixed-pressure case.

\subsection{nonzero terminated velocity}

When cyclotron cooling is absent
$(\sigma\subs{},\psiei,\epsilon\subs{})=(0.5,0.5,0)$,
the bremsstrahlung luminosity responses
for the nonzero terminated velocity condition
($\tau_0=0.01$, $\lambda_\tau=0$)
are almost the same as for the ``perfect stationary-wall'' condition
($\tau_0=0$, $\lambda_\tau=0$).
The corresponding phases are approximately equal
(to within $\LS0.005\pi$ radians,
see Table~\ref{'table.luminosity.altbc'}, bottom row).
The amplitudes agree to within a few percent.

When bremsstrahlung and cyclotron cooling are equally efficient at the shock
$(\sigma\subs{},\psiei,\epsilon\subs{})=(0.5,0.5,1)$,
zero- and nonzero-velocity termination
have different luminosity responses for the lowest three modes,
but approximately equal luminosity responses for the higher-order modes.
When cyclotron cooling dominates,
$(\sigma\subs{},\psiei,\epsilon\subs{})=(0.5,0.5,100)$,
the velocity termination affects the luminosity responses
for all the modes studied.

It is not clear why nonzero velocity termination at the lower boundary
affects luminosity responses of low-harmonic modes
more than higher modes.
Higher modes have $\lambda_{\rm br}$ profiles
with more ``nodes'' and ``antinodes'',
which should be expected to make them more sensitive to altered conditions
at many locations in the post-shock flow;
lower modes have fewer amplitude extrema 
and lesser spatial gradients of phase.
It can be speculated that for modes of higher harmonic number,
the integration in (\ref{'eq.def.luminosity'})
involves a proportionately smaller contribution
in the low-$\xi$,
compared to the case of lower modes.

\subsection{Comparison with numerical results}

In numerical studies of noise-driven shock oscillations,
in low-field (low $\epsilon\subs{}$) cases,
Wood \etal (1992) 
determined the lag of cyclotron luminosity response 
relative to the bremsstrahlung luminosity response,
which is comparable to the difference $\Phi_{\rm cy}-\Phi_{\rm br}$
in the notation of the present paper.
For cases with comparable equilibrium luminosities of 
cyclotron and bremsstrahlung ($L_{\rm cy,0}/L_{\rm br,0}=0.5$)
the cyclotron phase lag was $-0.8\pi$ radians 
for the fundamental ($n=1$) mode,
and $-0.2\pi$ for the first overtone ($n=2$).
This result is approximately consistent with
the typical lag $\sim-0.6\pi$ 
for the fundamental mode
in the fixed-density, fixed-temperature and fixed-velocity 
lower boudnary conditions for cases with $\epsilon\subs{}=1$
in Table~\ref{'table.luminosity.altbc'},
but inconsistent with fixed-flow-rate and fixed-pressure conditions.
Comparison of the cyclotron luminosity lags for the $n=2$ mode
may not be meaningful because 
in the analytic results,
the $n=2$ mode's phasing is sensitive to the system parameters
as well as the lower boundary condition. 
Wood \etal (1992) also caution that phasing may differ between 
free and driven oscillations;
by assumption,
our analytically represented oscillations are physically free.

As Imamura (1985) recognised,
the discrepent results from numerical investigations 
of radiative shock instabilities in accreting white dwarfs
may be largely affected by
the numerical treatment of the cold, high-density, gas 
near the lower boundary,
where rapid cooling causes
several of the fluid variables to experience sharp gradients.
In the Lagrangian numerical hydrodynamic calculations of
Imamura, Wolff \& Durisen (1984)
the grid zones accumulated at the lower boundary
were eliminated by coalescing the cells.
In Imamura (1985) the cluttered grid zones
were instead eliminated after cooling below a threshold temperature.
For the latter method the shock stabilities
were less sensitive to the procedural details
than in the previous method.
It is possible that methods involving mergers or eliminations
of material accreted at the base of the flow
have an effectively open lower boundary condition.
If the merger/deletion process still conserves
mass, energy and momentum 
then it still may effectively delete structural information
such as the ideally steep gradients of velocity and density near
the lower boundary.
The importance of those gradients is shown by the 
different oscillatory responses of the luminosities,
between the ``perfect stationary-wall'' 
and ''nonzero terminated velocity'' boundary conditions.

The Eulerian calculations
(\eg Langer, Chanmugam \& Shaviv 1981, 1982;
Chanmugam, Langer \& Shaviv 1985)
necessarily introduced a spatial filter to remove
noise and numerical instabilities.
This process may also affect the thermal instability
by smoothing away oscillatory structures,
such as those in the sensitive region 
immediately above the lower boundary,
that vary on finer spatial scales than the filtering resolution
(\eg nodes and phase jumps like those in the eigenfunctions
of higher modes in Saxton \& Wu 2001).

\begin{table}
\renewcommand{\baselinestretch}{1.0}
\caption{
Bremsstrahlung and cyclotron luminosity responses,
for the first six modes,
for various choices of the system parameters and lower boundary condition.
For the blocks of values from left to right, top to bottom,
the respective lower boundary conditions is
$\lambda_\pi=0$,
$\lambda_\zeta+\lambda_\tau=0$,
$\lambda_\zeta=0$,
$\lambda_\pi-\lambda_\zeta=0$,
($\lambda_\tau=0, \tau_0=0$)
and ($\lambda_\tau=0, \tau_0=0.01$).
The amplitudes are normalised by $\varepsilon$
and the phases are multiples of $\pi$ radians.
}
\label{'table.luminosity.altbc'}
{\tiny
\begin{center}
$
\begin{array}{ccc}
 &\lambda_\pi=0&\lambda_\zeta+\lambda_\tau=0\\
\begin{array}{rrr}
\sigma\subs{}&\psiei&\epsilon\subs{}\\
\\
\\
\hline
	\begin{array}{r}
	\\
	0.5\\ \\ \\ \\ \\ \\
	\\
	0.5\\ \\ \\ \\ \\ \\
	\\
	0.5\\ \\ \\ \\ \\ \\ 
	\\
	\end{array}&
	\begin{array}{r}
	\\
	0.5\\ \\ \\ \\ \\ \\ 
	\\
	0.5\\ \\ \\ \\ \\ \\ 
	\\
	0.5\\ \\ \\ \\ \\ \\ 
	\\
	\end{array}&
	\begin{array}{r}
	\\
	0\\ \\ \\ \\ \\ \\ 
	\\
	1\\ \\ \\ \\ \\ \\ 
	\\
	100\\ \\ \\ \\ \\ \\ 
	\\
	\end{array}
\end{array}
 &
 \begin{array}{cccc}
	\label{'table.luminosity.altbc.pi'}
	{{|L_{\rm br,1}|}\over{\varepsilon L_{\rm br,0}}}&\Phi_{\rm br}&
	{{|L_{\rm cy,1}|}\over{\varepsilon L_{\rm cy,0}}}&\Phi_{\rm cy}\\
	\hline
	\\
	\begin{array}{r}
	3.257\\9.652\\11.742\\12.355\\12.881\\15.137
	\end{array}&
	\begin{array}{r}
	0.347\\-0.968\\-0.336\\0.259\\0.962\\-0.444
	\end{array}&
	\begin{array}{r}
	.\\.\\.\\.\\.\\.
	\end{array}&
	\begin{array}{r}
	.\\.\\.\\.\\.\\.
	\end{array}
	\\
	\\
	\begin{array}{r}
	7.391\\8.167\\9.930\\11.913\\10.855\\14.405
	\end{array}&
	\begin{array}{r}
	0.485\\0.859\\-0.434\\0.062\\0.718\\-0.727
	\end{array}&
	\begin{array}{r}
	4.075\\0.521\\0.185\\0.344\\0.339\\0.329
	\end{array}&
	\begin{array}{r}
	-0.993\\0.825\\0.074\\-0.024\\-0.052\\-0.104
	\end{array}
	\\
	\\
	\begin{array}{r}
	24.160\\18.551\\7.654\\19.076\\15.504\\13.482
	\end{array}&
	\begin{array}{r}
	0.461\\0.485\\-0.752\\-0.519\\-0.224\\0.292
	\end{array}&
	\begin{array}{r}
	3.441\\0.474\\0.930\\1.164\\1.146\\1.088
	\end{array}&
	\begin{array}{r}
	0.987\\-0.657\\-0.208\\-0.231\\-0.216\\-0.191
	\end{array}
	\\
 \end{array}
 &
 \begin{array}{cccc}
	\label{'table.luminosity.altbc.zeta+tau'}
	{{|L_{\rm br,1}|}\over{\varepsilon L_{\rm br,0}}}&\Phi_{\rm br}&
	{{|L_{\rm cy,1}|}\over{\varepsilon L_{\rm cy,0}}}&\Phi_{\rm cy}\\
	\hline
	\\
	\begin{array}{r}
	18.303\\9.318\\7.515\\7.168\\6.567\\5.283
	\end{array}&
	\begin{array}{r}
	-0.622\\-0.015\\0.690\\-0.655\\-0.042\\0.634
	\end{array}&
	\begin{array}{r}
	.\\.\\.\\.\\.\\.
	\end{array}&
	\begin{array}{r}
	.\\.\\.\\.\\.\\.
	\end{array}
	\\
	\\
	\begin{array}{r}
	15.852\\7.924\\7.384\\5.919\\6.044\\4.845
	\end{array}&
	\begin{array}{r}
	-0.744\\-0.017\\0.564\\-0.808\\-0.218\\0.374
	\end{array}&
	\begin{array}{r}
	4.061\\0.506\\0.194\\0.343\\0.339\\0.331
	\end{array}&
	\begin{array}{r}
	-0.996\\0.823\\0.077\\-0.023\\-0.055\\-0.105
	\end{array}
	\\
	\\
	\begin{array}{r}
	17.381\\2.268\\4.668\\3.054\\3.195\\3.602
	\end{array}&
	\begin{array}{r}
	0.934\\-0.357\\0.156\\0.537\\-0.943\\-0.568
	\end{array}&
	\begin{array}{r}
	3.304\\0.420\\0.966\\1.174\\1.139\\1.077
	\end{array}&
	\begin{array}{r}
	0.967\\-0.578\\-0.202\\-0.229\\-0.211\\-0.189
	\end{array}
	\\
 \end{array}
\\
\hline
\\
\\
\\
 &\lambda_\zeta=0&\lambda_\pi-\lambda_\zeta=0\\
\begin{array}{rrr}
\sigma\subs{}&\psiei&\epsilon\subs{}\\
\\
\\
\hline
	\begin{array}{r}
	\\
	0.5\\ \\ \\ \\ \\ \\
	\\
	0.5\\ \\ \\ \\ \\ \\
	\\
	0.5\\ \\ \\ \\ \\ \\ 
	\\
	\end{array}&
	\begin{array}{r}
	\\
	0.5\\ \\ \\ \\ \\ \\ 
	\\
	0.5\\ \\ \\ \\ \\ \\ 
	\\
	0.5\\ \\ \\ \\ \\ \\ 
	\\
	\end{array}&
	\begin{array}{r}
	\\
	0\\ \\ \\ \\ \\ \\ 
	\\
	1\\ \\ \\ \\ \\ \\ 
	\\
	100\\ \\ \\ \\ \\ \\ 
	\\
	\end{array}
 \end{array}
 &
 \begin{array}{cccc}
 \label{'table.luminosity.altbc.zeta'}
	{{|L_{\rm br,1}|}\over{\varepsilon L_{\rm br,0}}}&\Phi_{\rm br}&
	{{|L_{\rm cy,1}|}\over{\varepsilon L_{\rm cy,0}}}&\Phi_{\rm cy}\\
	\hline
	\\
	\begin{array}{r}
	2.721\\1.657\\2.468\\1.247\\1.631\\1.946
	\end{array}&
	\begin{array}{r}
	-0.375\\-0.757\\-0.446\\0.134\\0.938\\-0.466
	\end{array}&
	\begin{array}{r}
	.\\.\\.\\.\\.\\.
	\end{array}&
	\begin{array}{r}
	.\\.\\.\\.\\.\\.
	\end{array}
	\\
	\\
	\begin{array}{r}
	1.272\\0.286\\1.103\\0.906\\1.093\\0.984
	\end{array}&
	\begin{array}{r}
	-0.355\\0.644\\-0.807\\-0.226\\0.429\\-0.944
	\end{array}&
	\begin{array}{r}
	4.079\\0.190\\0.513\\0.344\\0.339\\0.331
	\end{array}&
	\begin{array}{r}
	-0.995\\0.0077\\-0.825\\-0.024\\-0.054\\-0.105
	\end{array}
	\\
	\\
	\begin{array}{r}
	0.885\\0.124\\0.441\\0.397\\0.475\\0.643
	\end{array}&
	\begin{array}{r}
	-0.415\\-0.172\\0.787\\-0.887\\-0.351\\0.033
	\end{array}&
	\begin{array}{r}
	3.597\\0.429\\0.951\\1.173\\1.143\\1.080
	\end{array}&
	\begin{array}{r}
	0.976\\-0.585\\-0.203\\-0.229\\-0.211\\-0.189
	\end{array}
	\\
 \end{array}
 &
 \begin{array}{cccc}
 \label{'table.luminosity.altbc.pi-zeta'}
	{{|L_{\rm br,1}|}\over{\varepsilon L_{\rm br,0}}}&\Phi_{\rm br}&
	{{|L_{\rm cy,1}|}\over{\varepsilon L_{\rm cy,0}}}&\Phi_{\rm cy}\\
	\hline
	\\
	\begin{array}{r}
	2.729\\1.647\\2.452\\1.231\\1.613\\1.925
	\end{array}&
	\begin{array}{r}
	-0.376\\-0.755\\-0.477\\0.132\\0.937\\-0.467
	\end{array}&
	\begin{array}{r}
	.\\.\\.\\.\\.\\.
	\end{array}&
	\begin{array}{r}
	.\\.\\.\\.\\.\\.
	\end{array}
	\\
	\\
	\begin{array}{r}
	1.284\\0.277\\1.099\\0.897\\1.085\\0.969
	\end{array}&
	\begin{array}{r}
	-0.356\\0.635\\-0.811\\-0.232\\0.425\\-0.949
	\end{array}&
	\begin{array}{r}
	4.079\\0.513\\0.190\\0.344\\0.339\\0.331
	\end{array}&
	\begin{array}{r}
	-0.995\\0.825\\0.077\\-0.024\\-0.054\\-0.105
	\end{array}
	\\
	\\
	\begin{array}{r}
	0.923\\0.141\\0.443\\0.393\\0.457\\0.636
	\end{array}&
	\begin{array}{r}
	-0.419\\-0.223\\0.778\\-0.911\\-0.362\\0.023
	\end{array}&
	\begin{array}{r}
	3.598\\0.429\\0.951\\1.173\\1.143\\1.080
	\end{array}&
	\begin{array}{r}
	0.976\\-0.585\\-0.203\\-0.229\\-0.211\\-0.189
	\end{array}
	\\
 \end{array}
\\
\hline
\\
\\
\\
 &\lambda_\tau=0&\lambda_\tau=0, \tau_0=0.01\\
\begin{array}{rrr}
\sigma\subs{}&\psiei&\epsilon\subs{}\\
\\
\\
\hline
	\begin{array}{r}
	\\
	0.5\\ \\ \\ \\ \\ \\
	\\
	0.5\\ \\ \\ \\ \\ \\
	\\
	0.5\\ \\ \\ \\ \\ \\ 
	\\
	\end{array}&
	\begin{array}{r}
	\\
	0.5\\ \\ \\ \\ \\ \\ 
	\\
	0.5\\ \\ \\ \\ \\ \\ 
	\\
	0.5\\ \\ \\ \\ \\ \\ 
	\\
	\end{array}&
	\begin{array}{r}
	\\
	0\\ \\ \\ \\ \\ \\ 
	\\
	1\\ \\ \\ \\ \\ \\ 
	\\
	100\\ \\ \\ \\ \\ \\ 
	\\
	\end{array}
 \end{array}
 &
 \begin{array}{cccc}
 \label{'table.luminosity.altbc.tau'}
	{{|L_{\rm br,1}|}\over{\varepsilon L_{\rm br,0}}}&\Phi_{\rm br}&
	{{|L_{\rm cy,1}|}\over{\varepsilon L_{\rm cy,0}}}&\Phi_{\rm cy}\\
	\hline
	\\
	\begin{array}{r}
	2.745\\1.618\\2.497\\1.306\\1.641\\1.983
	\end{array}&
	\begin{array}{r}
	-0.373\\-0.766\\-0.458\\0.112\\0.911\\-0.490
	\end{array}&
	\begin{array}{r}
	.\\.\\.\\.\\.\\.
	\end{array}&
	\begin{array}{r}
	.\\.\\.\\.\\.\\.
	\end{array}
	\\
	\\
	\begin{array}{r}
	1.318\\0.345\\1.197\\1.015\\1.228\\1.109
	\end{array}&
	\begin{array}{r}
	-0.349\\0.615\\-0.820\\-0.240\\0.406\\-0.974
	\end{array}&
	\begin{array}{r}
	4.079\\0.513\\0.190\\0.344\\0.339\\0.331
	\end{array}&
	\begin{array}{r}
	-0.995\\0.825\\0.077\\-0.024\\-0.054\\-0.105
	\end{array}
	\\
	\\
	\begin{array}{r}
	1.254\\0.183\\0.715\\0.652\\0.804\\1.102
	\end{array}&
	\begin{array}{r}
	-0.415\\-0.047\\0.743\\-0.923\\-0.399\\-0.021
	\end{array}&
	\begin{array}{r}
	3.600\\0.429\\0.950\\1.173\\1.143\\1.080
	\end{array}&
	\begin{array}{r}
	0.976\\-0.585\\-0.202\\-0.229\\-0.211\\-0.189
	\end{array}
	\\
 \end{array}
 &
 \begin{array}{cccc}
 \label{'table.luminosity.altbc.tau.cut'}
	{{|L_{\rm br,1}|}\over{\varepsilon L_{\rm br,0}}}&\Phi_{\rm br}&
	{{|L_{\rm cy,1}|}\over{\varepsilon L_{\rm cy,0}}}&\Phi_{\rm cy}\\
	\hline
	\\
	\begin{array}{r}
	2.718\\1.723\\2.664\\1.451\\1.797\\2.139
	\end{array}&
	\begin{array}{r}
	-0.363\\-0.782\\-0.454\\0.122\\0.909\\-0.490
	\end{array}&
	\begin{array}{r}
	.\\.\\.\\.\\.\\.
	\end{array}&
	\begin{array}{r}
	.\\.\\.\\.\\.\\.
	\end{array}
	\\
	\\
	\begin{array}{r}
	1.196\\1.193\\0.414\\1.020\\1.238\\1.115
	\end{array}&
	\begin{array}{r}
	-0.324\\-0.806\\0.641\\-0.215\\0.424\\-0.958
	\end{array}&
	\begin{array}{r}
	4.085\\0.190\\0.514\\0.343\\0.341\\0.331
	\end{array}&
	\begin{array}{r}
	-0.995\\0.082\\0.823\\-0.024\\-0.054\\-0.104
	\end{array}
	\\
	\\
	\begin{array}{r}
	0.445\\0.047\\0.211\\0.170\\0.202\\0.257
	\end{array}&
	\begin{array}{r}
	-0.364\\-0.098\\0.844\\-0.821\\-0.288\\0.081
	\end{array}&
	\begin{array}{r}
	3.768\\0.411\\0.966\\1.169\\1.124\\1.069
	\end{array}&
	\begin{array}{r}
	0.971\\-0.595\\-0.188\\-0.229\\-0.215\\-0.193
	\end{array}
	\\
 \end{array}
\\
\hline
\end{array}
$
\end{center}
}
\end{table}

\section{Conclusions}
\label{'section.conclusions'}

The oscillatory and stability properties
are not significantly affected
for modifications of the stationary-wall boundary condition
involving the perturbed hydrodynamic variables:
for constant pressure, flow rate, density or temperature
at the lower boundary.
The cases considered here have eigenvalues
which are effectively identical to the ``perfect stationary-wall'' results,
or at most the ``small offset'' profiles
which only affect the $\lambda$-functions 
is regions very close to the lower boundary.

If the stationary solution terminates at a lower boundary with nonzero velocity,
the oscillatory frequencies are essentially unchanged
from the ``perfect stationary-wall'' case.
However the nonzero-terminated velocity modes are slightly more stable
than the corresponding modes of the ``perfect stationary-wall,''
and the higher modes are more greatly affected.

The cyclotron luminosity response
is independent of the lower boundary condition
as this emission is peaked near the shock,
where jump conditions specify the hydrodynamic variables
and the lower boundary conditions have little influence.
The bremsstrahlung luminosity response
is a sensitive probe of the conditions at the lower boundary,
because bremsstrahlung emission is peaked near that boundary
and the perturbed hydrodynamic variables in this region
are locally very sensitive to changes in the lower boundary conditions.
However the fixed-density and fixed-temperature lower boundary conditions
have modes with identical luminosity responses;
the reason for this equivalence is unclear.

Termination of the stationary solution at nonzero velocity
produces effectively identical luminosity responses
as the same modes with zero velocity termination,
for all modes when cyclotron cooling is unimportant
(\ie when $\epsilon\subs{}$ is small).
As $\epsilon\subs{}$ increases,
the velocity termination makes a difference
to the luminosity responses of the lowest-order modes
but not higher-order modes.
When $\epsilon\subs{}$ is sufficiently large,
nonzero velocity termination affects
the luminosity responses of all modes studied.
For modest two-temperature effects,
the bremsstrahlung luminosity responses of higher-order modes
tend (with few exceptions) to be
greater under the fixed-pressure lower boundary condition,
and lower under the fixed flow-rate lower boundary condition.
However there is no universal pattern between $\delta$
and the luminosity responses for general $(\sigma\subs{},\psiei)$.
For a given mode and given system parameters,
the bremsstrahlung responses in the fixed-pressure and fixed-flow cases
are approximately in antiphase.

Past numerical simulations reproduce some of what may be
the important qualitative features of 
these thermally unstable radiative shocks,
though it is not clear which, if any,
of the numerical treatments provides superior predictions
about the stability properties and frequencies of the shock oscillations.
All the  treatments to date
may lack sufficient spatial and temporal resolution
in the lower boundary region, 
where the oscillatory instability may originate.
Indeed treatments may alter the fundamental character of the boundary,
(\eg the removal of grid zones or accreted material
may provide a more open boundary 
than would be realistic for the fixed wall of the white dwarf surface).

\section*{Acknowledgements}

I thank Kinwah~Wu, James~Imamura
and the anonymous referee for thorough comments on the manuscript,
which have contributed greatly to the paper's improvement and
clarification.

\section*{References}






\reference Beardmore,~A.P. \& Osborne,~J.P., 1997, MNRAS 286, 77
\reference Bertschinger,~E. 1986, ApJ 304, 154
\reference Chanmugam,~G., Langer,~S.H. \& Shaviv,~G., 1985, ApJ 299, L87 
\reference Chevalier,~R.A. \& Imamura,~J.N 1982 ApJ 261, 543
\reference Cropper,~M. 1990 Sp. Sci. Rev., 54, 195
\reference Field,~G.B., 1965, ApJ, 142, 531
\reference Imamura,~J.N., 1985, ApJ, 296, 128
\reference Imamura,~J.N., Aboasha,~A., Wolff,~M.T. \& Wood,~K.S. 1996 ApJ, 458, 327
\reference Imamura,~J.N. \& Steiman-Cameron,~T.Y., 1986, ApJ 311, 786
\reference Imamura,~J.N., Wolff,~M.T., \& Durisen,~R.H., 1984, ApJ, 276, 667
\reference Langer,~S.H., Chanmugam,~G. \& Shaviv,~G. 1981 ApJ 245, L23
\reference Langer,~S.H., Chanmugam,~G. \& Shaviv,~G. 1982 ApJ 258, 289
\reference Larsson,~S. 1987, A\&A 181, L15
\reference Larsson,~S. 1989, A\&A 217, 146
\reference Melrose,~D.B. 1986, Instabilities in Laboratory and Space Plasmas, Cambridge University Press, Cambridge
\reference Middleditch,~J., Imamura,~J.N. \& Steiman-Cameron,~T.Y., 1987,
ApJ 489, 912
\reference Ramsayer,~T.F., Robinson,~E.L., Zhang,~E., Wood,~J.H.
\& Stiening,~R.F. 1993, MNRAS 260, 209
\reference Saxton,~C.J. 1999 PhD Thesis, University of Sydney
\reference Saxton,~C.J. \& Wu,~K. 1999 MNRAS 310, 677
\reference Saxton,~C.J. \& Wu,~K. 2001 MNRAS 324, 659
\reference Saxton,~C.J., Wu,~K. \& Pongracic,~H., 1997, PASA 14, 164
\reference Saxton,~C.J., Wu,~K., Pongracic,~H. \& Shaviv,~G., 1998, MNRAS 299, 862
\reference Spitzer,~L. 1962, Physics of Fully Ionized Gases, New York: Interscience, 135
\reference T\'{o}th,~G. \& Draine,~B.T. 1993 ApJ, 413, 176
\reference Wolff,~M.T., Gardner,~J. \& Wood,~K.S. 1989, ApJ 346, 833
\reference Wolff,~M.T., Wood,~K.S. \& Imamura,~J.N. 1991, ApJ 375, L31
\reference Wolff,~M.T., Imamura,~J.N., Middleditch,~J., Wood,~K.S.
\& Steiman-Cameron,~T. 1999, 
Annapolis Workshop on Magnetic Cataclysmic 
Variables, ASP Conf. Ser. 157, ed. Hellier,~C. \& Mukai,~K., 149
\reference Wood,~K.S., Imamura,~J.N., \& Wolff,~M.T. 1992, ApJ 398, 593
\reference Wu,~K. 1994 PASA, 11, 61
\reference Wu, K., 2000, Space Sci. Rev., 93, 611

\end{document}